\documentclass[12pt]{article}
\usepackage[dvips]{graphicx}
\usepackage{feynmf}

\begin{document}


\begin{titlepage}
\noindent
\begin{flushright}
hep-ph/0111003 \\
UCD-2001-09\\
DESY 01-180 \\
\end{flushright}
\begin{center}
  \begin{Large}
    \begin{bf}
	$D$-term challenges for supersymmetric \\ 
          gauged abelian flavor symmetries 
    \end{bf}
  \end{Large}
\end{center}
\vspace{0.2cm}
\begin{center}
  Brandon Murakami$^a$, Kazuhiro Tobe$^a$, and James D. Wells$^{a,b}$\\
  \vspace{0.2cm}
  \begin{it}
        
   ${}^{(a)}$Davis Institute for High Energy Physics, Department of Physics\\
	University of California, One Shields Avenue, Davis, CA 95616 \\
   ${}^{(b)}$Deutsches Elektronen-Synchrotron DESY, D-22603 Hamburg, Germany \\
  \end{it}

\end{center}

\begin{abstract}
Attention is called to potentially dangerous lepton-flavor violation (LFV)
induced by the $D$-terms of additional $U(1)$ flavor-dependent gauge
symmetries in supersymmetric models.  In such models, LFV persists 
despite an arbitrarily high scale for the $U(1)$ breaking and despite
arbitrarily small gauge couplings.  In light of recent experimental
observations of neutrino oscillations, these models are highly motivated
experimentally and theoretically.  Taking into account also the recent
measurement of the muon anomalous magnetic moment,
strong bounds are calculated for
the magnitude of the $D$-term-induced LFV. Using current data we find that 
the slepton mass-mixing parameter
$(m_{\tilde l_L}^2)_{12}/m_{\tilde l_L}^2$ is required to be less than
$\mathcal{O}(10^{-4})$ --- a value perhaps already 
too low compared to expectations. Near future
probes will increase sensitivity to $10^{-6}$.

\end{abstract}
\end{titlepage}

\section{Motivation}
%
Recent atmospheric~\cite{SuperK_atm} and solar~\cite{Solar} neutrino
experiments provide convincing evidence for physics beyond the
standard model (SM). The interesting points are not only that they
suggest non-zero neutrino masses, but also that they indicate the existence of
large flavor mixings in the lepton sector. The different flavor
structure between quark and lepton
sectors would be an important hint for the fermion mass hierarchy problem.
One interesting approach to accommodate both large mixing in the lepton
sector and small mixing in the quark sector is the introduction of a
$U(1)$ flavor-dependent symmetry.  So far many models with the
$U(1)$ flavor symmetry have been proposed to explain the
observed fermion masses and mixings~\cite{neutrino_models1}. Such a $U(1)$
symmetry may originate from string theory~\cite{string}.

Another interesting possible indication of new physics beyond the SM
is the recent result for the muon anomalous magnetic moment (muon $g$--2) 
by the E821 experiment at Brookhaven National Laboratory~\cite{E821}.
It is found that the muon $g$--2 measurement is $2.6 \sigma$ away from the
SM prediction\footnote{QCD contribution to muon $g$--2 in the SM
prediction is still under debate~\cite{SM_g2}.}:
\begin{eqnarray}
a_\mu({\rm exp}) - a_\mu({\rm SM}) =43(16)\times 10^{-10}.
\label{E821_result}
\end{eqnarray}
Since the size of the deviation is of the same order as the electroweak
contribution to muon $g$--2, the result implies 
new physics around the TeV scale~\cite{muon_th}. 

Weak scale supersymmetric (SUSY) extensions of the SM, which are well 
motivated by the hierarchy problem, provide a natural explanation 
of the anomaly of muon 
$g$--2~\cite{muon_susy} when superpartners are light.
In the SUSY version of the seesaw mechanism~\cite{seesaw},
non-zero neutrino masses are also naturally accommodated. 
For the present experimental status, the SUSY models are
the best motivated extensions of the SM.
 
In general, if there are extra $U(1)$ gauge symmetries present in a
supersymmetric model, the breaking of these symmetries will induce $D$-term
contributions to the  scalar masses~\cite{Dterm}.  Of the many 
interesting models with extra $U(1)$ gauge symmetries, we choose to focus on
$U(1)$ flavor symmetries due to the recent intriguing neutrino
observations.  In the proposed models with $U(1)$ flavor
symmetry~\cite{neutrino_models1}, large flavor mixings may
exist in the lepton sector. Thus the $D$-term contributions may induce
large lepton flavor violation (LFV) through the sleptons. 
Since relatively light sleptons are expected to explain the anomaly
of muon $g$--2, the sleptons cannot decouple to suppress the large LFV
from the $D$-term contributions. 
Therefore, experimental bounds on LFV processes strongly 
constrain SUSY models with $U(1)$ flavor symmetry.

\section{$l_i \rightarrow l_j \gamma$ 
and muon $g$--2 correlations}
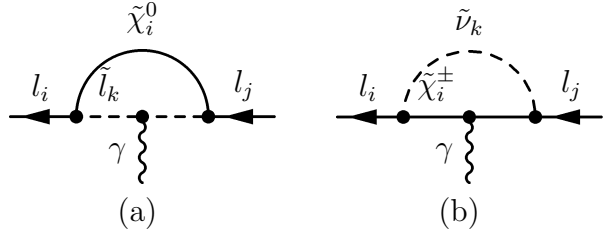
\begin{figure}[t]
\center
\begin{tabular}{cc}
\begin{fmffile}{fig-neutralino}
\begin{fmfgraph*}(100,50)
\fmfleft{mu}
\fmfright{e}
\fmfbottom{ext}
\fmf{fermion,label=$l_j$}{e,v1}
\fmf{dashes}{v1,v3}
\fmf{dashes,label=$\tilde l_k$}{v3,v2}
\fmf{fermion,label=$\l_i$}{v2,mu}
\fmffreeze
\fmf{plain,right,label=$\tilde\chi_i^0$}{v1,v2}
\fmf{photon,label=$\gamma$}{v3,ext}
\fmfdot{v1,v2,v3}
\end{fmfgraph*}
\end{fmffile}
&
\begin{fmffile}{fig-chargino}
\begin{fmfgraph*}(100,50)
\fmfleft{mu}
\fmfright{e}
\fmfbottom{ext}
\fmf{fermion,label=$l_j$}{e,v1}
\fmf{plain}{v1,v2}
\fmf{plain,label=$\tilde\chi^\pm_i$}{v2,v3}
\fmf{fermion,label=$\l_i$}{v3,mu}
\fmffreeze
\fmf{dashes,right,label=$\tilde\nu_k$}{v1,v3}
\fmf{photon,label=$\gamma$}{v2,ext}
\fmfdot{v1,v2,v3}
\end{fmfgraph*}
\end{fmffile}\\
(a) & (b) \\
\end{tabular}

\caption{At 1-loop level, $\mu \to e \gamma$, $\tau\to\mu\gamma$, and
the muon anomalous magnetic moment occur through neutralino and
chargino loops.}
\label{fig.diagrams}
\end{figure}

%
Here we briefly demonstrate constraints on the LFV slepton masses from
the $\mu \rightarrow e \gamma$ and $\tau \rightarrow \mu \gamma$ processes
in light of the recent muon $g$--2 result~\cite{HT}.
Processes $\mu \rightarrow e \gamma$ and $\tau \rightarrow \mu \gamma$
are generated by one-loop diagrams mediated by sleptons, neutralinos, and 
charginos in the presence of LFV in the slepton masses
(Fig.~\ref{fig.diagrams}).
When the left-handed sleptons have a LFV mass between the first
and second generations $(m^2_{\tilde{l}_L})_{12}$,
$\mu \rightarrow e \gamma$ process is induced. The branching ratio is 
given by
\begin{eqnarray}
{\rm Br}(\mu \rightarrow e \gamma) &\simeq& 
\frac{\pi}{75}\alpha \left(\alpha_2+\frac{5}{4} \alpha_Y
\right)^2 \frac{\tan^2\beta}{G_{\rm F}^2 m_{\rm SUSY}^4}
\left(\frac{(m^2_{\tilde{l}_L})_{12}}{m^2_{\rm SUSY}}
\right)^2,
\label{br_meg}
\end{eqnarray}
where, for illustrative purposes, we simply assumed that all SUSY 
mass parameters are the same scale $m_{\rm SUSY}$.
From the current experimental limit
Br$(\mu \rightarrow e \gamma)<1.2 \times 10^{-11}$ (MEGA)~\cite{meg_exp},
we get a constraint on the LFV slepton mass:
\begin{eqnarray}
\frac{(m^2_{\tilde{l}_L})_{12}}{m^2_{\rm SUSY}}
< 3 \times 10^{-4} \left(\frac{30}{\tan \beta} \right)
\left(\frac{m_{\rm SUSY}}{400~{\rm GeV}} \right)^2.
\label{LFV_limit}
\end{eqnarray}
Note that if SUSY particles get heavier, the limit becomes
weaker. Therefore, one solution to suppress the large LFV would be
a ``decoupling solution'' in which the first and second generation
sleptons are heavy enough to avoid the LFV problem~\cite{decoupling}.

However, when considering the muon $g$--2 measurement, the limit in 
Eq.~(\ref{LFV_limit}) becomes much stricter. Since the $\mu \rightarrow 
e \gamma$ process originates from the same type of diagrams as
muon $g$--2 (but without the slepton mass-mixing) (Fig.~\ref{fig.diagrams}), 
the branching ratio of
$\mu \rightarrow e \gamma$ and the SUSY contribution to muon $g$--2
are correlated,\footnote{
In muon $g$--2, SUSY contribution from charginos and left-handed sleptons
loop tend to be dominant. Therefore, if only right-handed sleptons
have LFV masses, the correlation between Br$(\mu \rightarrow e \gamma)$
and $\delta a_\mu^{\rm SUSY}$ becomes weaker (Ref.~\cite{HT}).
}
as stressed in Ref.~\cite{HT}. If we take the soft mass 
parameters $M_1$, $M_2$, $m_{\tilde l}$, and $\mu$ to be equal at the
weak scale, the SUSY contribution to $a_\mu$ is given by
\begin{eqnarray}
\delta a_\mu^{\rm SUSY} &\simeq& \frac{5 \alpha_2 + \alpha_Y}{48 \pi}
\frac{m_\mu^2}{m_{\rm SUSY}^2} \tan \beta,\\
&=& 24\times 10^{-10} \left( \frac{\tan\beta}{30} \right) 
\left( \frac{400~{\rm GeV}}{m_{\rm SUSY}} \right)^2.
\label{SUSY_amu}
\end{eqnarray}
Note that a relatively light SUSY scale of the order 
$\mathcal{O}$(100 GeV)
can explain the E821 experiment in Eq.~(\ref{E821_result}).

From Eqs.~(\ref{br_meg}) and (\ref{SUSY_amu}), we can obtain a relation
between Br$(\mu \rightarrow e \gamma)$ and $\delta a_\mu^{\rm SUSY}$. Then
taking into account the limit on Br$(\mu \rightarrow e \gamma)$ and
$\delta a_\mu^{\rm SUSY}$ ($\delta a_\mu^{\rm SUSY} >10^{-9}$ at $2 \sigma$),
we obtain a constraint on the LFV mass in terms of observables \cite{HT}:
\begin{eqnarray}
\frac{(m^2_{\tilde{l}_L})_{12}}{m^2_{\rm SUSY}}
<7 \times 10^{-4} \left(\frac{10^{-9}}{\delta a_\mu^{\rm SUSY}}
\right) \left( \frac{{\rm Br}(\mu \rightarrow e \gamma)}{1.2\times 10^{-11}}
\right)^{1/2}.
\label{LFV_limit2}
\end{eqnarray}
Comparing this result with that of Eq.~(\ref{LFV_limit}), we should
note an important difference.  That is, we cannot have the decoupling
solution if $\delta a_\mu^{\rm SUSY}$ is fixed.  Similarly, from $\tau
\rightarrow \mu \gamma$, we have a bound on the LFV mass
$(m^2_{\tilde{L}})_{23}$ from the present experimental limit Br$(\tau
\rightarrow \mu \gamma)<1\times 10^{-6}$~\cite{tmg_exp}:
\begin{eqnarray}
\frac{(m^2_{\tilde{l}_L})_{23}}{m^2_{\rm SUSY}}
<5 \times 10^{-1} \left(\frac{10^{-9}}{\delta a_\mu^{\rm SUSY}}
\right) \left( \frac{{\rm Br}(\tau \rightarrow \mu \gamma)}{1\times 10^{-6}}
\right)^{1/2}.
\end{eqnarray}

These findings signify an important correlation between the muon
$g$--2 result and the problem of SUSY LFV.
Therefore, the search for LFV processes will be significantly
sensitive to any origins of LFV slepton masses.
So far it has been pointed out that 
high-energy flavor-violating interactions~\cite{HKR}, such as
GUT interactions~\cite{GUT_LFV} and
large neutrino Yukawa interactions~\cite{neutrino_LFV}, can 
induce significant LFV in the slepton masses~\cite{review_LFV}.
In the next section, we will show that $D$-term contributions
of a $U(1)$ vector multiplet may generate large LFV in the slepton
masses, and therefore searches for LFV severely constrain SUSY models
with extra $U(1)$ gauge symmetries.

\section{$D$-terms and sfermion mass-mixing}
Although $D$-terms are 
a generic feature of extra $U(1)$ gauge symmetries in SUSY
models, the Froggatt-Nielsen mechanism~\cite{FN} is chosen to
illustrate their effects on sfermion masses and mass-mixings
while also addressing the fermion mass hierarchy problem for the
quarks and leptons.  In this letter, we consider only one $U(1)$ flavor
symmetry denoted $U(1)_F$.  However in the more general case of
multiple $U(1)$ vector multiplets, there will simply be an additive
effect of the $D$-terms. In this framework, Yukawa couplings originate
from the following operator:
\begin{eqnarray}
W_{\rm Yukawa} = f_{ij} \left(\frac{\phi}{M_*} \right)^{Q_i+Q_j}
\psi^c_i \psi_j H~({\rm{or}}~\bar{H}).
\label{yukawa_op}
\end{eqnarray}
Here $M_*$ is a fundamental scale of the theory.
Field $\psi_i$ represents ordinary quarks and leptons, whose $U(1)_F$
charge is $Q_i$ ($Q_i\geq 0$ by construction). 
Here we assumed that the $U(1)_F$ charges
of the Higgs 
fields ($H$ and $\bar{H}$) and a field $\phi$ are $0$ and $-1$, 
respectively. After the
$U(1)_F$ symmetry is broken, the $\phi$ field gets a vacuum expectation value (vev)
with $\langle \phi \rangle < M_*$, generating
the hierarchical Yukawa couplings.

In SUSY models, if the $U(1)_F$ is a local symmetry, the breaking 
induces $D$-term contributions to the scalar masses.
If the $U(1)_F$ symmetry is broken by fields $\phi_\pm$ of charge $\pm1$,
the $D$-term ($D_F$) obtains a vev\footnote{Explicit models for the
$U(1)_F$ breaking can be found in Ref.~\cite{Dterm}.}:
\begin{eqnarray}
\langle D_F \rangle = g_F \left( \langle \phi_+ \rangle^2
-\langle \phi_-\rangle^2 \right)\simeq -\frac{1}{g_F}(m_+^2-m_-^2).
\end{eqnarray}
Here $m_\pm$ are SUSY breaking masses for the $\phi_\pm$ fields, and $g_F$ is
the $U(1)_F$ gauge coupling.
The non-zero vev for the $D$-term gives contributions to the squark
and slepton masses~\cite{Dterm}:
\begin{eqnarray}
{\cal{L}_{\rm mass}} = -\sum_i Q_i \Delta m^2 {\tilde \psi}^*_i 
{\tilde{\psi}}_i,
\label{scalar_mass}
\end{eqnarray}
where $\Delta m^2=m^2_- - m^2_+$. Different Yukawa interactions between
$\phi_\pm$ induce non-zero $\Delta m^2$ through the renormalization
group (RG) running from the fundamental scale $M_*$ to 
the $U(1)_F$ breaking
scale $M_F$ ($M_F\sim \langle \phi_\pm \rangle$) even if $m_+^2 =
m_-^2$ is assumed at $M_*$. Large Yukawa couplings are expected
to radiatively break the $U(1)_F$; therefore, $\Delta
m^2$ is expected to be of order (weak scale)$^2$.

In general, a higher theory may present the fermions in the gauge
interaction basis to be different from the mass basis, as seen in
Eq.~(\ref{yukawa_op}).  After rotating to the mass eigenstates 
($\psi_i'=V^\dagger_{ij} \psi_j$) for the
fermions, the sfermion mass terms in Eq.~({\ref{scalar_mass}) get
flavor mixings:
\begin{eqnarray}
{\cal L}_{\rm mass} = -\sum_{ijk} V^*_{k i} Q_k \Delta m^2 V_{kj}
\tilde{\psi}_i^{*'} \tilde{\psi}_j'.
\label{D_LFV}
\end{eqnarray}
If the mixing $V_{ij}$ is large, the $D$-term contributions
may generate large flavor violating sfermion masses provided
$Q_k$ are not all the same value.  Therefore
Eq.~(\ref{D_LFV}) embodies two strong statements that may affect large 
classes of model building:  The scalar masses induced by the $D$-term
does not explicitly depend on the gauge coupling $g_F$ nor the
$U(1)_F$ breaking scale $M_F$.  This feature is not limited the
Froggatt-Nielsen mechanism.  That is, this effect persists for any
SUSY model of $U(1)$ flavor-dependent gauge bosons with
\emph{arbitrarily high} mass and \emph{arbitrarily small} gauge
couplings.\footnote{
If there is an additional $U(1)$ flavor-independent gauge symmetry, there
is a mechanism to suppress the $U(1)_F$ $D$-term contributions along the
lines of Ref.~\cite{babu}, which introduces an additional $U(1)$ flavor
symmetry group.  In our context, to effectively banish all possible
problematic $D$-terms, the additional symmetry should be flavor-independent
with much larger gauge coupling than the original flavor symmetry.
}

In the models~\cite{neutrino_models1} motivated by recent
neutrino results, large mixings exist in the lepton sector.
For example, in Ref.~\cite{SY}, lopsided $U(1)_F$ charges 
to the left-handed lepton
doublets $l_i$ $(i=1-3)$, namely $Q_{l_2}=Q_{l_3}=0$ and $Q_{l_1}=+1$,
naturally explain the large mixing for atmospheric neutrinos
(and solar neutrinos if the solar neutrino solution is the large mixing angle 
solution (LMA))~\cite{SY}. In this model, a large mixing can be induced in 
the left-handed slepton masses due to the $D$-term contribution in 
Eq.~(\ref{D_LFV}) :
\begin{eqnarray}
\frac{(m^2_{\tilde{l}_L})_{12}}{m^2_{\rm SUSY}}
=V^*_{11} \frac{\Delta m^2}{m^2_{\rm SUSY}} V_{12}
\end{eqnarray}
If the solar neutrino solution is the LMA solution (which is the most favored
solar neutrino solution by SuperKamiokande and SNO experiments at present), 
$|V_{11}V_{12}|$ can be as large as $0.1$~\cite{SY}. 
Therefore this $D$-term contribution to the left-handed sleptons can
be very large. For example, if $\Delta m^2/m_{\rm SUSY}^2$ is as low
as $1/16\pi^2$ and $|V_{11}V_{12}|$ as low as $0.01$, we would still get
the large result $(m^2_{\tilde{l}_L})_{12}/m^2_{\rm SUSY}\simeq 10^{-4}$.

In this context,
the ``anarchy''~\cite{anarchy} 
type $U(1)$ charge assignment ($Q_{l_1}=Q_{l_2}=Q_{l_3}$) may be an
interesting solution to avoid large LFV as well as to explain
the large mixing for neutrinos.  These types of models predict the LMA
solution for the solar neutrino problem and large ``reactor angle''
$U_{e3}$.  Therefore future neutrino experiments will probe such models.

\begin{figure}[t]
\center
\resizebox{5 in}{!}{\includegraphics{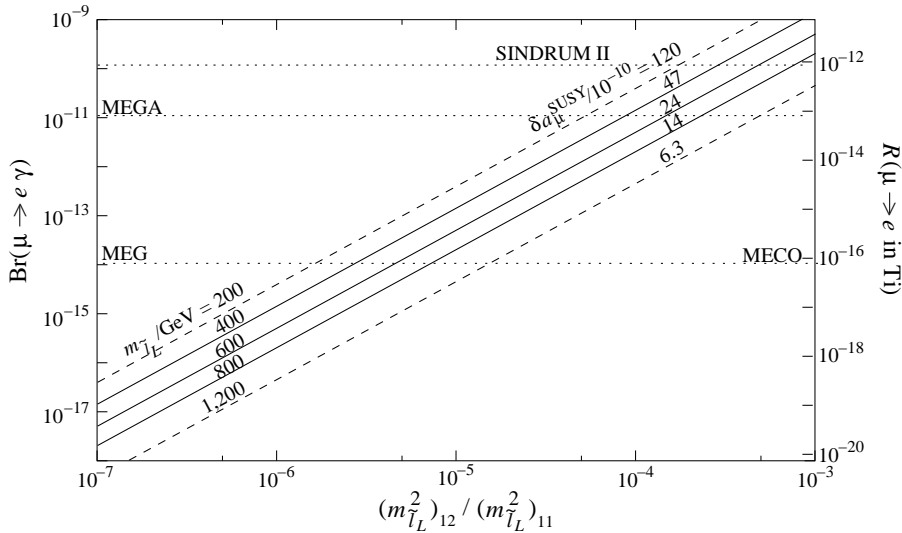}}
\caption{The slepton mass-mixing magnitude vs. the $\mu \to e \gamma$
branching ratio (left axis) and the muon conversion ratio $R=\sigma(\mu N
\to eN)/\sigma(\mu N \to \nu N')$ for titanium (right axis).  The
parameters $M_2 = 125$ GeV, $\tan\beta=30$, $\mu>0$, and $A_0=0$ are
applied to the mSUGRA scenario.  The extra $U(1)$ is broken at $10^{15}$
GeV.  
The value of the left-handed
slepton mass parameter is varied for the diagonal lines.
The horizontal lines are the current (MEGA and SINDRUM II)
and near future (MEG and MECO) experimental limits from $\mu \to e
\gamma$ and $\mu \rightarrow e$ conversion experiments.  
The future experiments MEG and MECO will both approximately equally place the
strongest experimental limits on the slepton mass-mixing.} 
\label{fig-constraints}
\end{figure}

In Fig.~\ref{fig-constraints}, the $\mu \rightarrow e \gamma$
branching ratio and $\mu \rightarrow e$ conversion rate are shown
as a function of the slepton-mixing
$(m^2_{\tilde{l}_L})_{12}/(m^2_{\tilde{l}_L})_{11}$ assumed to originate from
a $U(1)_F$ $D$-term.  Here the universal soft SUSY breaking 
parameters are
defined at the GUT scale. 
The RG
equations for the couplings and masses are numerically solved, and at
the weak scale we calculate the event rates for $\mu \rightarrow e
\gamma$ and $\mu \rightarrow e$ conversion in Ti. We take
$\tan\beta=30$, $\mu>0$, $M_2=125$ GeV at the weak scale, and $A_0=0$
at the GUT scale, and vary the slepton mass $m_{\tilde{l}_L}$ at the
weak scale.  We assume that $U(1)_F$ breaking scale $M_F$ 
is $10^{15}$ GeV. 
In general, the $U(1)_F$
gauge interaction can induce the flavor violating effect through
the RG running from GUT scale to $M_F$. However, here we neglect 
the RG effect of the $U(1)_F$ gauge coupling, and hence
our calculated event rates will be conservative\footnote{
We have checked that our event rates do not strongly depend 
on the $U(1)_F$ breaking scale $M_F$.
However, if the scale $M_F$ is close to TeV scale, the extra $U(1)_F$
gauge boson also contributes to the LFV 
processes~\cite{zprim}, and the event rates will be increased.}.
Here we also assume that only the left-handed sleptons have the LFV masses.
If there are flavor mixings in the right-handed slepton sector,
the branching ratio will be increased. Thus our calculated rates
will be very conservative.

In Fig.~\ref{fig-constraints}, we also show the SUSY contribution
to the muon $g$--2 observable $\delta a_\mu^{\rm SUSY}$.
As can be seen from the figure, the present $\mu \rightarrow e \gamma$
null results from MEGA strongly constrain the LFV mass
$(m^2_{\tilde{l}_L})_{12}$
in the region where the present muon $g$--2 result is favored:
$(m^2_{\tilde{l}_L})_{12}/(m^2_{\tilde{l}_L})_{11} \leq 10^{-4}$.
This detailed
analysis confirms the naive estimate of the constraint on the LFV
masses in the previous section. Therefore, many models with the gauged $U(1)$
flavor symmetry are significantly constrained.

In SUSY models with slepton mixings, the photon penguin diagram 
tends to dominate $\mu \rightarrow eee$
and $\mu \rightarrow e$ conversion in nuclei. Thus the following
relations amongst the event rates are approximately held:

\begin{eqnarray}
\frac
{R(\mu \rightarrow e~{\rm in~Ti~(Al)})} 
{{\rm Br}(\mu \rightarrow e \gamma)}
&\simeq& 5~(3) \times 10^{-3},
\\
\frac
{{\rm Br}(\mu \rightarrow eee)}
{{\rm Br}(\mu \rightarrow e \gamma)}
&\simeq& 6 \times 10^{-3}.
\end{eqnarray}
\begin{table}[t]
\small
\begin{tabular}{|c|c|c||c|}
\hline
Process & Current limit & Proposed & 
Further possibility \\
 & &sensitivity & \scriptsize{(PRISM, NUFACT)}\\
\hline \hline
$\mu \rightarrow e \gamma$ & $1.2\times 10^{-11}$ \footnotesize{(MEGA)}& 
$2\times 10^{-14}$ \footnotesize{(MEG)} 
& $\sim 10^{-15}$\\
\hline
$\mu N \rightarrow e N$ & $6.1 \times 10^{-13}$ \scriptsize{(SINDRUM II)} & 
$5\times 10^{-17}$ \footnotesize{(MECO)}
& $\sim 10^{-18}$\\
\hline
$\mu \rightarrow eee$ & $1.0\times 10^{-12}$ \scriptsize{(SINDRUM)}
& --- & $\sim 10^{-16}$\\
\hline
\end{tabular}
\caption{Current limits and proposed sensitivities for event rates
of muon flavor violating processes.}
\label{table}
\end{table}
As shown in Table~\ref{table},
the proposed experiments MEG at PSI~\cite{PSI} for $\mu \rightarrow e
\gamma$ and MECO at BNL~\cite{MECO} for $\mu\rightarrow e$ conversion
will increase sensitivities of the event rates to $10^{-14}$
and $10^{-16}$ respectively, and hence they
will probe nearly two orders of magnitude past the current
$(m^2_{\tilde{l}_L})_{12}/(m^2_{\tilde{l}_L})_{11}$ limit of
Eq.~(\ref{LFV_limit2}) (Fig.~{\ref{fig-constraints}}).  
Further distant experiments that utilize
intense sources of low energy muons (PRISM and NuFACT)~\cite{review_LFV, PRISM}
will probe to nearly three orders of magnitude over the current
limit.   Such a bound of
$(m^2_{\tilde{l}_L})_{12}/(m^2_{\tilde{l}_L})_{11}$ less than nearly
$10^{-7}$ has potential to greatly change our theoretical perspective.
These probes of LFV therefore warrant great attention.

The $\tau \rightarrow \mu \gamma$ process is also important since
it can independently put a constraint on the other LFV mass.
Although the present constraint is not very strong, the effort
to improve the sensitivity of $\tau \rightarrow \mu \gamma$ will
be very important.

In light of the recent muon $g$--2 result, we have shown that 
LFV searches are quite
sensitive to the $D$-term contributions induced by a $U(1)$ flavor
symmetry. Already the present experiments strongly constrain
SUSY models with $U(1)$ flavor symmetry. The future LFV experiments
as well as neutrino and muon $g$--2 experiments will either find signals
of flavor violation or force us to reevaluate what general approaches
to the theory of flavor are viable.

\section*{Acknowledgements}
We thank K.S. Babu for informing us about Ref.~{\cite{babu}}.
This work was funded in part by the Department of Energy, the Alfred P. Sloan
Foundation, and DESY.

\end{document}